\documentclass[a4paper,UKenglish,cleveref, autoref, thm-restate]{oasics-v2021}

\pdfoutput=1 
\hideOASIcs 

\usepackage{algorithm}
\usepackage[noend]{algpseudocode}

\title{An Evaluation of the State-of-the-Art Software and Hardware Implementations of BIKE} 

\author{Andrea Galimberti\footnote{Corresponding author}}
{Dipartimento di Elettronica, Informazione e Bioingegneria (DEIB), Politecnico di Milano,
Piazza Leonardo da Vinci 32, 20133 Milano, Italy}
{andrea.galimberti@polimi.it}
{https://orcid.org/0000-0003-0254-3933}
{}

\author{Gabriele Montanaro}
{Dipartimento di Elettronica, Informazione e Bioingegneria (DEIB), Politecnico di Milano,
Piazza Leonardo da Vinci 32, 20133 Milano, Italy}
{gabriele.montanaro@polimi.it}
{https://orcid.org/0000-0003-1119-2629}
{}

\author{William Fornaciari}
{Dipartimento di Elettronica, Informazione e Bioingegneria (DEIB), Politecnico di Milano,
Piazza Leonardo da Vinci 32, 20133 Milano, Italy \and \url{https://fornaciari.faculty.polimi.it/}}
{william.fornaciari@polimi.it}
{https://orcid.org/0000-0001-8294-730X}
{}

\author{Davide Zoni}
{Dipartimento di Elettronica, Informazione e Bioingegneria (DEIB), Politecnico di Milano,
Piazza Leonardo da Vinci 32, 20133 Milano, Italy \and \url{https://zoni.faculty.polimi.it/}}
{davide.zoni@polimi.it}
{https://orcid.org/0000-0002-9951-062X}
{}

\authorrunning{A. Galimberti, G. Montanaro, W. Fornaciari and D. Zoni} 

\Copyright{Andrea Galimberti, Gabriele Montanaro, William Fornaciari and Davide Zoni} 


\category{} 

\relatedversion{} 

\funding{This work was partially supported by SIAE MICROELETTRONICA and
	by the EU Horizon 2020 “TEXTAROSSA” project (Grant No. 956831).}

\nolinenumbers 

\begin{document}

\maketitle

\begin{abstract}
NIST is conducting a process for the standardization of post-quantum cryptosystems,
i.e., cryptosystems that are resistant to attacks by both traditional and quantum computers
and that can thus substitute the traditional public-key cryptography solutions which are
expected to be broken by quantum computers in the next decades.
This manuscript provides an overview and a comparison of the existing state-of-the-art
implementations of the BIKE QC-MDPC code-based post-quantum KEM, a candidate in
NIST's PQC standardization process.
We consider both software, hardware, and mixed hardware-software implementations
and evaluate their performance and, for hardware ones, their resource utilization.
\end{abstract}

\section{Introduction}
\label{sec:intro}
Traditional public-key cryptosystems~(PKC), including RSA~\cite{Rivest_CommACM1978},
ECDSA~\cite{Bernstein_PKC2006}, and Diffie-Hellman~\cite{Diffie_TIT1976}, underpin
cryptographically secure key exchange mechanisms and digital signature schemes.
Such cryptoschemes are however expected to be broken by quantum computers in the
upcoming decades~\cite{NSA_CNSA2.0}.
The threat posed by quantum computers requires the definition and the design of alternative
cryptosystems that perform the same functions as PKC ones, maintaining security against
traditional computer attacks while ensuring security against quantum computer attacks.
Post-quantum cryptography (PQC) aims to develop cryptosystems that are resistant to both
traditional attacks and new quantum attack models, which can be implemented on traditional
architecture computers and existing devices, and that can be integrated into the networks
and communication protocols currently in use~\cite{Bernstein_Nature2017}.

The National Institute of Standards and Technology~(NIST) is conducting a process for
the standardization of PQC cryptosystems, in particular key encapsulation mechanisms~(KEM)
and digital signature schemes~\cite{NIST_website}.
After the third round of the PQC standardization process, NIST selected the CRYSTALS-Kyber
lattice-based KEM for standardization while appointing the fourth evaluation round to analyze
further the code-based BIKE, Classic McEliece, and HQC and the isogeny-based SIKE.
The performance of both the software and hardware implementations of such cryptosystems is
crucial for evaluating the cryptosystems, in addition to security against traditional and quantum attacks.
In particular, NIST takes Intel Haswell processors and Xilinx Artix-7 FPGAs as
reference platforms for software and hardware implementations, respectively.

A KEM allows the secure transmission, through a public key algorithm, of a shared secret
that can then be expanded to generate keys to be used in a symmetric cryptosystem,
which is more efficient than a PKC scheme for the transmission of long messages~\cite{Shoup_IACR2001}.
After generating a random element of the finite group that underlies
the implemented public key scheme, this element is exchanged between
the two communicating parties, which can finally derive the shared secret
by applying a hash function to the element of the finite group.

BIKE is a post-quantum KEM based on quasi-cyclic moderate-density parity-check~(QC-MDPC) codes~\cite{BIKE_website}.
These codes are used in a scheme similar to that first proposed by Niederreiter~\cite{Niederreiter_PCIT1986}.
BIKE distinguishes itself for its good trade-off between ciphertext and key lengths and performance,
making it a good candidate for standardization after the fourth round~\cite{NIST_IR8413}.
Instances of BIKE are specified for NIST security levels 1, 3, and 5, providing security
against quantum attacks equivalent to AES-128, -192, and -256, respectively.

\begin{algorithm}[t]
	\caption{Primitives of the BIKE key encapsulation mechanism~\cite{BIKE_website}.}
	\label{alg:bike}
	\begin{algorithmic}[1]
		\Function{$[H, \sigma, h]$ KeyGen\ }{$seed$, $\sigma$}
			\State $H = [h_0 | h_1] = $ PRNG$(seed);$
			\State $h = h_1 \odot h_{0}^{-1};$
			\State\Return $\{H, h, \sigma\};$
		\EndFunction
		\Function{$[K, c]$ Encaps\ }{$h$, $m$}
			\State $e = $ \textbf{H}$(m);$
			\State $s = e_0 \oplus (e_1 \odot h);$
			\State $m' = m \ \oplus $ \textbf{L}$(e);$ 
			\State $K = $\textbf{K}$(\{m,\{s, m'\}\});$
			\State\Return $\{K, \{s, m'\}\};$
		\EndFunction
		\Function{$[K]$ Decaps\ }{$H$, $\sigma$, $c$}
			\State $e' = $ BGFD\scriptsize{ECODER} \normalsize $(s, H);$
			\State $m'' = m' \ \oplus $ \textbf{L}$(e');$
			\State $a = (e' == $ \textbf{H}$(m''))$ ? $m''$ : $\sigma;$
			\State$K = $ \textbf{K}$(\{a, c\});$
			\State\Return $K;$
		\EndFunction
	\end{algorithmic}
\end{algorithm}

The BIKE cryptosystem can be split into three primitives.
Key generation produces a private-public key pair~(\textsc{KeyGen} in Algorithm~\ref{alg:bike}),
encapsulation generates a shared secret and encrypts it with the public key~(\textsc{Encaps}),
and decapsulation retrieves the shared secret with the private key from the ciphertext~(\textsc{Decaps}).
Due to its QC-MDPC code-based nature, BIKE uses binary polynomial
arithmetic operations and the Black-Gray-Flip decoding procedure~\cite{Drucker_PQCrypto2020},
while random number generation and cryptographic hash functionalities~(\textbf{H},
\textbf{K}, and \textbf{L} in Algorithm~\ref{alg:bike}) are implemented by employing
SHA-3 and SHAKE.

\subsection*{Contributions}
This manuscript provides an overview and a comparison of the existing state-of-art
implementations of BIKE, a QC-MDPC code-based post-quantum KEM candidate for
standardization in the fourth round of NIST's PQC standardization process.

The goal is to gauge the ability to deploy BIKE on different computing platforms
suitable to various real-world use-case scenarios, ranging from
low-power embedded systems to desktop-class CPUs and mid-range FPGAs.

\section{Related works}
\label{sec:soa}
The literature contains a variety of proposals that provide complete software 
and hardware implementations of QC-MDPC code-based post-quantum cryptosystems.

\subsection{State-of-the-art software implementations}
On the software side, implementations of QC-MDPC code-based cryptosystems participating
in the NIST PQC competition were made publicly available and distributed open-source.

Two separate software versions of LEDAcrypt, an early candidate to the NIST's PQC standardization process
which was admitted to its third round of evaluation, are available at \cite{LEDAcrypt_website}.
They consist of a reference version written in plain C11 and an optimized one that exploits
the AVX2 extension for recent Intel Core CPUs.

\cite{BIKE_website} provides instead the two official software implementations of BIKE, a reference one written
in plain C11 and an optimized one that exploits the Intel AVX512 extension.
Other works from literature provide software implementations for ISAs other than the Intel x86 one,
with \cite{Chen_TCHES2021} targeting Arm Cortex-M4 microcontrollers and \cite{Chen_ACNS2022}
introducing support for RISC-V computing platforms.
Further additional implementations of BIKE, including
a fully portable one,
versions optimized for AVX2 and AVX512 instruction set extensions, and
implementations optimized for CPUs that support \emph{PCLMULQDQ} and \emph{VPCLMULQDQ} instructions,
are also publicly available on Github~\cite{BIKE_SW_Github}.

The Intel AVX2 instruction set extension and similar ones can indeed significantly
boost the performance of binary polynomial arithmetic operations.
Intel introduced the \emph{PCLMULQDQ} instruction and the corresponding hardware support
in its Westmere architecture to accelerate the
AES Galois/Counter Mode~(AES-GCM) authenticated encryption algorithm.
The \emph{PCLMULQDQ} instruction performs the carry-less multiplication of two 64-bit
operands.
Similarly, the ARMv8-A architecture provides the \emph{VMULL.P64} instruction,
which takes as inputs two 64-bit NEON registers and outputs their product,
computing according to binary polynomial multiplication, on a 128-bit NEON register.

The work in \cite{Drucker_ARITH2018} leveraged the \emph{VPCLMULQDQ} instruction,
which is intended to further accelerate AES-GCM and which is the vectorized extension
of \emph{PCLMULQDQ}, to compute multiplications between large-degree binary polynomials,
i.e.. polynomials with degree greater than 511.
\cite{Drucker_CSCML2020} introduced a constant-time algorithm for
polynomial inversion, targeting the software implementation of BIKE
and based on Fermat's little theorem.
The authors optimized the exponentiation operation and further improved performance
by means of a source code targeting the latest Intel Ice Lake CPUs, that support
the AVX512 and \emph{VPCLMULQDQ} instructions.
The optimizations introduced in \cite{Drucker_ARITH2018} and \cite{Drucker_CSCML2020}
are implemented within the Intel AVX2-optimized constant-time implementations of BIKE~\cite{BIKE_SW_Github}.

\subsection{State-of-the-art hardware and hardware-software implementations}
On the hardware side, the literature provides a variety of FPGA-based implementations of
QC-MDPC code-based cryptosystems.

\cite{Heyse_CHES2013, Maurich_DATE2014} proposed the implementation of
the McEliece cryptosystem with QC-MDPC codes on FPGAs.
In particular, \cite{Heyse_CHES2013} targeted a performance-oriented design
while \cite{Maurich_DATE2014} focused on a resource-optimized one. 
\cite{Hu_TC2017} discussed a fast implementation of QC-MDPC Niederreiter encryption for
FPGAs, outperforming the work in \cite{Heyse_CHES2013} thanks to using a hardware module
to estimate the Hamming weight of large vectors and proposing a hardware implementation
tailored to low-area devices for encryption and decryption used in QC-MDPC code-based cryptosystems.

The authors of BIKE presented a VHDL FPGA-based implementation, targeting Xilinx Artix-7 FPGAs
and providing support for the key generation, encryption, and decryption KEM primitives
on a unique design~\cite{Richter-Brockmann_TC2021}.
However, the proposed architecture was custom-tailored to smaller FPGA targets,
up to Artix-7 100, and it employed the AES and SHA-2 cryptographic functions
as random oracles, thus supporting a now obsolete specification of BIKE.
The work in \cite{Richter-Brockmann_TC2021} provided the first FPGA-based implementation of
the BGF decoder, employed a multiplication module that minimized the BRAM usage by parallelizing the
computation of a simpler schoolbook multiplication algorithm, rather than applying
a more complex one such as Karatsuba's, and implemented binary polynomial inversion
by employing a Fermat-based inversion algorithm that is a variant of the algorithm
introduced in \cite{Hu_TCSII2015}.

The work in \cite{Galimberti_DSD2022} presented another FPGA-based implementation of BIKE,
split into two components devoted to supporting the client-side~(key generation and decapsulation)
and server-side~(encapsulation) primitives.
The client and server cores integrated highly configurable hardware accelerators
for binary polynomial multiplication~\cite{Barenghi_ICECS2019,Zoni_Access2020_Mul}
and inversion~\cite{Galimberti_TC2022} and BGF decoding~\cite{Zoni_Access2020_Dec}.
Setting different parameters for the configurable accelerators allowed the authors to implement
the client and server cores on FPGAs ranging from Artix-7 35 to Artix-7 200.

Finally, \cite{Richter-Brockmann_TCHES2022} proposed an updated FPGA-based implementation
of \cite{Richter-Brockmann_TC2021} that employed a Keccak core rather than AES and SHA-2 ones,
as specified in the latest version of the BIKE cryptoscheme~\cite{BIKE_spec4.2}.
In addition, the work in \cite{Richter-Brockmann_TCHES2022} only implements
a dense-sparse multiplication module, which exploits the sparse representation
of one of the two operands in the binary polynomial multiplication, rather than
also a dense-dense one, and implements the extended Euclidean algorithm for
binary polynomial inversion rather than the Fermat-based one.
The proposed architecture targets Artix-7 FPGAs and the authors listed
three instances implementing the whole KEM providing a range of area-performance trade-offs.
The smallest one requires less resources than the lightweight one from \cite{Richter-Brockmann_TC2021}
and provides a more than 3$\times$ speedup, while the largest one takes 3.7ms compared to
the 4.8ms of the high-speed one from \cite{Richter-Brockmann_TC2021} while also
occupying a smaller area.

\smallskip
On the hardware-software~(HW/SW) side,
\cite{Montanaro_ICECS2022} proposed a mixed HW/SW approach that
made use of three HLS-generated accelerators, each implementing one of the BIKE primitives.
The HW/SW approach allowed mixing the usage of hardware acceleration for the most
computationally expensive primitives with the software execution of the least complex ones.
The proposed solution resulted in different combinations of hardware-implemented
and software-executed KEM primitives on three chips from the Xilinx Zynq-7000 heterogeneous
SoC family, which feature ARM CPUs coupled with programmable FPGA logic equivalent
to the Artix-7 one.

\section{Methodology}
\label{sec:meth}
The evaluation of state-of-the-art BIKE implementations spanned software,
hardware, and hardware-software ones.
On the software side, it considered 32- and 64-bit architectures, ARM and x86 ISAs,
embedded- and desktop-class processors, and plain-C and AVX2-optimized software.
On the hardware and hardware-software sides, we compared solutions that were
human-designed and HLS-generated, targeting Xilinx FPGAs and heterogeneous SoCs.

\subsection{Evaluated software implementations}
\label{ssec:setup_sw}
The software performance analysis considered three implementations of BIKE.

\smallskip\noindent
The \textbf{reference C99}~(\textbf{Ref C99} in Section~\ref{sec:exp}) software~\cite{BIKE_website}
is the reference implementation from the official BIKE NIST submission and provides a code without
any architecture-specific optimization, making it suitable to any target computing platform.

\smallskip\noindent
The \textbf{additional portable C99}~(\textbf{CT C99}) software~\cite{BIKE_SW_Github}, written in plain C99
without any architecture-specific optimization, delivers a constant-time execution and is compatible with
both 64-bit Intel and ARM architectures. 

\smallskip\noindent
The \textbf{additional Intel AVX2-optimized}~(\textbf{CT AVX2}) software~\cite{BIKE_SW_Github}
provides a faster constant-time implementation on Intel x86-64 CPUs that support the Intel AVX2
instruction set extension, i.e., CPUs from the Intel Haswell generation and later ones.

\subsection{Evaluated hardware and hardware-software implementations}
\label{ssec:setup_hw}
The experimental evaluation of hardware and hardware-software solutions considered
three different implementations of the BIKE cryptoscheme.

\smallskip\noindent
The \textbf{Official} hardware implementation~\cite{Richter-Brockmann_TCHES2022}
delivers a unified design that implements the whole BIKE KEM and executes it in
constant time.
The authors provide three instances ranging from a lightweight one up to mid-range
and high-performance ones.
The proposed design, targeting Xilinx FPGAs, is described in SystemVerilog and
publicly available online~\cite{RacingBIKE_Github}.

\smallskip\noindent
The \textbf{Client-server} hardware implementation~\cite{Galimberti_DSD2022}
consists of two separate architectures devoted to client- (key generation and decapsulation)
and server-side (encapsulation) operations of BIKE.
The two client and server cores integrate configurable components, whose selection of
the different architectural parameters results in instances targeting smaller and larger FPGAs.

\smallskip\noindent
The \textbf{HLS-based} hardware-software implementation~\cite{Montanaro_ICECS2022}
consists of three instances, targeting heterogeneous SoCs, that mix software execution
and hardware acceleration, through HLS-generated components, of the BIKE KEM primitives.
The instances differ in which primitives are executed in software and in hardware,
allowing them to fit on different target chips.

\subsection{Target computing platforms}
\label{ssec:setup_target}
The software implementations were executed on target platforms ranging
from low-end ARM-based embedded systems to desktop-class Intel CPUs, while
the hardware and hardware-software ones targeted Xilinx Artix-7 FPGAs and
Zynq-7000 SoCs, respectively.

\smallskip\noindent\textbf{Arm Cortex-A9}~(\textbf{ARM32} in Section~\ref{sec:exp})
is an embedded-class 32-bit processor implementing the ARMv7-A instruction set architecture~(ISA).
We execute BIKE on an Arm Cortex-A9 dual-core processor featured on a Xilinx Zynq-7000
heterogeneous SoC.
The ARM CPU has a clock frequency up to 667MHz, and the external memory
mounted on the employed Digilent Zedboard development board, which houses
the Zynq-7000 chip, is a 512MB DDR3.
The BIKE software is executed on top of the Xilinx Petalinux operating system.

\smallskip\noindent\textbf{Arm Cortex-A53}~(\textbf{ARM64})
is an embedded-class 64-bit processor implementing the ARMv8-A ISA.
In particular, we consider the RP3A0 system-in-package mounted on a Raspberry Pi Zero 2 W,
that features a quad-core 64-bit Arm Cortex-A53 processor clocked up to 1GHz and 512MB of SDRAM.
We executed BIKE on the Raspberry Pi running the 64-bit Raspberry Pi OS Lite operating system,
that is based on Debian 11, and setting a fixed 1GHz clock frequency through Linux \textit{cpupower} tools.

\smallskip\noindent\textbf{Intel Core i5-10310U}~(\textbf{Intel x86-64}) is a desktop-class
64-bit processor implementing the x86-64 ISA and providing support for the Intel AVX2 extension,
running at a clock frequency up to 4.4GHz.
The PC mounting the Intel CPU ran the Ubuntu 20.04.3 LTS operating system.
Such CPU supports the execution of the AVX2-optimized software version of BIKE.

\begin{table}[t]
	\centering
	\caption
	[Available FPGA resources on Xilinx Artix-7 FPGAs and Zynq-7000 SoCs.]
	{Available FPGA resources on FPGAs from the Xilinx Artix-7 family and SoCs from the Xilinx Zynq-7000 family.
		Legend: \textbf{LUT} look-up tables, \textbf{FF} flip-flops, \textbf{BRAM} 36kb blocks of block RAM, \textbf{DSP} digital signal processing slices.}
	\scalebox{1}{
		\begin{tabular}{lrrrr}
			\textbf{FPGA/SoC} & \textbf{LUT} & \textbf{FF} & \textbf{BRAM} & \textbf{DSP} \\ \hline
			Artix-7 12        &         8000 &       16000 &            20 &           40 \\
			Artix-7 15        &        10400 &       20800 &            25 &           45 \\
			Artix-7 25        &        14600 &       29200 &            45 &           80 \\
			Artix-7 35        &        20800 &       41600 &            50 &           90 \\
			Artix-7 50        &        32600 &       65200 &            75 &          120 \\
			Artix-7 75        &        47200 &       94400 &           105 &          180 \\
			Artix-7 100       &        63400 &      126800 &           135 &          240 \\
			Artix-7 200       &       134600 &      269200 &           365 &          740 \\ \hline
			Zynq-7000 Z-7010  &        17600 &       35200 &            60 &           80 \\
			Zynq-7000 Z-7015  &        46200 &       92400 &            95 &          160 \\
			Zynq-7000 Z-7020  &        53200 &      106400 &           140 &          220 \\ \hline
		\end{tabular}}
	\label{tab:avail_res}
\end{table}

\smallskip\noindent\textbf{Xilinx Artix-7}~(\textbf{A7-xxx}) FPGAs are mid-range, cost-effective FPGA chips
which are the suggested target for hardware implementations within the NIST PQC
standardization process, which targets FPGAs in order to prevent the adoption
of ASIC-specific technology optimizations and thus ensure a fair comparison of
the hardware implementations.
The look-up table~(LUT), flip-flop~(FF), block RAM~(BRAM), and digital signal processing~(DSP)
resources available on each FPGA chip from the Xilinx Artix-7 family are listed in Table~\ref{tab:avail_res}.

\smallskip\noindent\textbf{Xilinx Zynq-7000}~(\textbf{Z-70xx}) chips are heterogeneous SoCs
that couple an Arm Cortex-A9 dual-core processor with Artix-7 class
programmable FPGA logic.
The ARM CPU part has a clock frequency up to 667MHz, and the external memory
mounted on the employed Digilent Zedboard development board, which houses
the Zynq-7000 chip, is a 512MB DDR3.
The LUT, FF, BRAM, and DSP resources available on the considered Zynq-7000 SoCs
are listed in Table~\ref{tab:avail_res}.
The BIKE software~\cite{BIKE_website} is executed on top of the Xilinx Petalinux operating system
and extended with calls to the HLS-generated hardware accelerators.

\section{Experimental evaluation}
\label{sec:exp}
The experimental evaluation of the state-of-the-art implementations of BIKE considers
first the software solutions and then the hardware and hardware-software ones.
The discussion of the collected software performance results is split into
the absolute execution times, to gauge the actual real-world performance of the BIKE cryptoscheme,
and the relative execution times, to highlight similarities and differences between
the various computing platforms and software implementations.
The evaluation of the hardware state-of-the-art solutions is split instead into
their performance, expressed as their absolute execution time, and their FPGA
resource utilization, expressed in terms of LUT, FF, BRAM, and DSP resources.

\begin{table}[t]
	\centering
	\caption
	[Breakdown of the software execution times of BIKE.]
	{Breakdown of the execution times of BIKE, expressed in milliseconds,
		for different security levels, architectures, and software implementations.
		Legend: \textsc{KeyGen} key generation, \textsc{Encaps} encapsulation, \textsc{Decaps} decapsulation,
		\textbf{SL\textit{i}} NIST security level \textit{i}.}
	\scalebox{1}{
		\begin{tabular}{lrrrrrrrr}
			                       &                                        \multicolumn{8}{c}{\textbf{Target CPU, software version, security level}}                                        \\
			                       &  \multicolumn{2}{c}{\textbf{ARM32}}  & \multicolumn{2}{c}{\textbf{ARM64}}  &                 \multicolumn{4}{c}{\textbf{Intel x86-64}}                  \\
			                       & \multicolumn{2}{c}{\textbf{Ref C99}} & \multicolumn{2}{c}{\textbf{CT C99}} & \multicolumn{2}{c}{\textbf{CT C99}} & \multicolumn{2}{c}{\textbf{CT AVX2}} \\
			\textbf{KEM primitive} & \textbf{SL1} &          \textbf{SL3} & \textbf{SL1} &         \textbf{SL3} & \textbf{SL1} &         \textbf{SL3} & \textbf{SL1} &          \textbf{SL3} \\ \hline
			\textsc{KeyGen}        &       332.34 &                920.93 &        21.15 &                66.97 &         3.68 &                11.91 &         0.20 &                  0.57 \\
			\textsc{Encaps}        &        14.83 &                 40.94 &         1.99 &                 5.60 &         0.27 &                 0.77 &         0.05 &                  0.09 \\
			\textsc{Decaps}        &       464.82 &               1188.27 &        33.93 &               104.65 &         4.07 &                12.67 &         0.81 &                  2.55 \\ \hline
			\textbf{Overall KEM}   &       811.98 &               2150.14 &        57.06 &               177.23 &         8.02 &                25.35 &         1.06 &                  3.21 \\ \hline
		\end{tabular}
	}
	\label{tab:perf_sw}
\end{table}

\subsection{Software performance}
\label{ssec:abs_perf_sw}
The range of computing platforms and software implementations considered in the
experimental evaluation resulted in significant differences in terms of absolute performance
when executing the BIKE software, as shown by data provided in Table~\ref{tab:perf_sw}.
Such performance results were collected by executing BIKE 100 times and averaging
the ensuing execution times for each considered CPU and software version.

On the lower end, the \textbf{ARM32} 32-bit Arm Cortex-A9 platform, running at 667MHz,
provided execution times of 812ms and 2150ms, i.e., in the order of seconds,
when executing the \textbf{Ref C99} reference implementation
with NIST security levels 1 and 3, respectively.

Moving to a more efficient code that made use of 64-bit instructions, i.e.,
the \textbf{CT C99} additional portable implementation,
as well as to a more modern ARMv8-A architecture, provided a speedup of more than 10$\times$.
The performance on the \textbf{ARM64} Arm Cortex-A53 64-bit CPU, also running at a higher
1GHz clock frequency, measured at 57ms and 177ms for AES-128 and -192 security
instances of BIKE, respectively.

Executing the same \textbf{CT C99} software implementation of BIKE on the \textbf{Intel x86-64}
CPU resulted in a further speedup of around 7$\times$.
The different architecture and the higher clock frequency, in the order of 4GHz,
allowed executing BIKE instances with security levels 1 and 3 in 8ms and 25ms, respectively.

Finally, we evaluated the execution, on the same \textbf{Intel x86-64} CPU,
of the \textbf{CT AVX2} software implementation making use of instructions from
the Intel AVX2 extension.
The execution times of 1.1ms and 3.2ms are around 8$\times$ smaller than
those obtained by the \textbf{CT C99} plain-C99 software, which highlights
the effectiveness of those dedicated instructions in a software making wide
use of binary polynomial arithmetic.

\begin{table}[t]
	\centering
	\caption
	[Breakdown of the percentage execution times of BIKE for different security levels,
	architectures, and software implementations.]
	{Breakdown of the percentage execution times of BIKE for different security levels,
		architectures, and software implementations.
		Legend: \textsc{KeyGen} key generation, \textsc{Encaps} encapsulation, \textsc{Decaps} decapsulation,
				\textbf{SL\textit{i}} NIST security level \textit{i}.}
	\scalebox{1}{
		\begin{tabular}{llrrrrrrrr}
			                       &                     &                                        \multicolumn{8}{c}{\textbf{Target CPU, software version, security level}}                                        \\
			                       &                     &  \multicolumn{2}{c}{\textbf{ARM32}}  & \multicolumn{2}{c}{\textbf{ARM64}}  &                 \multicolumn{4}{c}{\textbf{Intel x86-64}}                  \\
			                       &                     & \multicolumn{2}{c}{\textbf{Ref C99}} & \multicolumn{2}{c}{\textbf{CT C99}} & \multicolumn{2}{c}{\textbf{CT C99}} & \multicolumn{2}{c}{\textbf{CT AVX2}} \\
			\textbf{KEM primitive} & \textbf{Operation}  & \textbf{SL1} &          \textbf{SL3} & \textbf{SL1} &         \textbf{SL3} & \textbf{SL1} &         \textbf{SL3} & \textbf{SL1} &          \textbf{SL3} \\ \hline
			\textsc{KeyGen}        & PRNG                &          0\% &                   0\% &          1\% &                  1\% &          0\% &                  0\% &          1\% &                   1\% \\
			                       & Inversion           &         39\% &                  41\% &         34\% &                 35\% &         43\% &                 44\% &         17\% &                  17\% \\
			                       & Multiplication      &          2\% &                   2\% &          2\% &                  2\% &          2\% &                  2\% &          1\% &                   1\% \\ \cline{3-10}
			                       &                     &         41\% &                  43\% &         37\% &                 38\% &         46\% &                 47\% &         19\% &                  18\% \\ \hline
			\textsc{Encaps}        & \textbf{H} function &          0\% &                   0\% &          1\% &                  1\% &          1\% &                  1\% &          2\% &                   1\% \\
			                       & Multiplication      &          2\% &                   2\% &          2\% &                  2\% &          2\% &                  2\% &          1\% &                   1\% \\
			                       & \textbf{L} function &          0\% &                   0\% &          0\% &                  0\% &          0\% &                  0\% &          1\% &                   1\% \\
			                       & \textbf{K} function &          0\% &                   0\% &          0\% &                  0\% &          0\% &                  0\% &          1\% &                   0\% \\ \cline{3-10}
			                       &                     &          2\% &                   2\% &          3\% &                  3\% &          3\% &                  3\% &          5\% &                   3\% \\ \hline
			\textsc{Decaps}        & Decoding            &         57\% &                  55\% &         56\% &                 56\% &         49\% &                 48\% &         71\% &                  75\% \\
			                       & \textbf{L} function &          0\% &                   0\% &          0\% &                  0\% &          0\% &                  0\% &          1\% &                   1\% \\
			                       & \textbf{H} function &          0\% &                   0\% &          1\% &                  1\% &          1\% &                  1\% &          1\% &                   1\% \\
			                       & \textbf{K} function &          0\% &                   0\% &          0\% &                  0\% &          0\% &                  0\% &          1\% &                   0\% \\ \cline{3-10}
			                       &                     &         57\% &                  55\% &         59\% &                 59\% &         51\% &                 50\% &         76\% &                  79\% \\ \hline
		\end{tabular}
	}
	\label{tab:perf_sw_ratio}
\end{table}

\subsection{Software performance profile}
\label{ssec:rel_perf_sw}
Table~\ref{tab:perf_sw_ratio} details the performance profile of
the software execution of BIKE, on the different computing platforms,
highlighting the ratio of execution time taken by the main operations
comprising the three primitives of the BIKE KEM.

On the \textbf{ARM32} ARMv7-A platform, the execution of the \textbf{Ref C99} reference
implementation resulted in a performance profile characterized by
binary polynomial inversion and BGF decoding occupying up to 41\% and 57\% of
the KEM execution time, with binary polynomial multiplication taking instead up to 4\% overall.

The execution of the \textbf{CT C99} additional portable implementation of BIKE
on the \textbf{ARM64} ARMv8-A CPU highlighted binary polynomial inversion and BGF decoding taking
up to 35\% and 56\% of the execution time.

Executing the same \textbf{CT C99} software on
the \textbf{Intel x86-64} processor saw the KEM execution time being
almost equally distributed between inversion and decoding, taking up to
44\% and 49\%, respectively.
Overall, the results are quite similar to ARMv8-A software execution,
due to not using any Intel-specific optimization.

On the contrary, the execution of the \textbf{CT AVX2} AVX2-optimized
software on the same \textbf{Intel x86-64} CPU produced quite different results.
The decoding procedure takes indeed a larger portion of the KEM execution time,
up to 75\%, while inversion only takes up to 17\%.
Notably, AVX2 instructions provide the higher speedup to the
operations in binary polynomial arithmetic, namely multiplications
and inversions, where the latter is computed as iterated multiplications 
and exponentiations.
Binary polynomial multiplications and inversions end up therefore taking
smaller shares of the KEM execution time.

Overall, the obtained results highlight QC-MDPC bit-flipping decoding
and binary polynomial inversion as the two operations taking the largest share
of the execution time across all considered platforms and software versions,
with an aggregate share of the execution time ranging from 89\% to 96\%.
The third largest share of execution time is occupied by binary
polynomial multiplications, ranging from 2\% to 4\%.
\textbf{H}, \textbf{K}, and \textbf{L} functions, which are not accelerated by AVX instructions,
require a notable share of execution time, 8\% and 5\% for NIST security levels 1 and 3,
respectively, only when executing AVX2-optimized software.

\begin{table}[t]
	\centering
	\caption
	[Breakdown of the hardware execution times of BIKE.]
	{Breakdown of the execution times of AES-128 security instances of BIKE,
		expressed in milliseconds, for different state-of-the-art FPGA-based implementations.
		Legend: \textbf{LW} lightweight, \textbf{MR} mid-range, \textbf{HP} high-performance instances,
		\textbf{*} aggregate for key generation and decapsulation.}
	\scalebox{1}{
		\begin{tabular}{lrrrrrrrr}
			                       &                                      \multicolumn{8}{c}{\textbf{Hardware implementation}}                                      \\
			                       &  \multicolumn{3}{c}{\textbf{Official}}  & \multicolumn{2}{c}{\textbf{Client-server}} & \multicolumn{3}{c}{\textbf{HLS-based}}  \\
			\textbf{KEM primitive} & \textbf{LW} & \textbf{MR} & \textbf{HP} &    \textbf{LW} &               \textbf{HP} & \textbf{LW} & \textbf{MR} & \textbf{HP} \\ \hline
			\textsc{KeyGen}        &        3.79 &        1.87 &        1.67 & \textbf{*}5.71 &            \textbf{*}0.58 &      137.84 &      332.14 &      137.84 \\
			\textsc{Encaps}        &        0.44 &        0.28 &        0.13 &           0.03 &                      0.03 &       14.86 &       14.86 &       14.86 \\
			\textsc{Decaps}        &        6.90 &        4.21 &        1.89 & \textbf{*}5.71 &            \textbf{*}0.58 &      464.61 &      135.48 &      135.48 \\ \hline
			\textbf{Overall KEM}   &       11.14 &        6.36 &        3.70 &           5.74 &                      0.61 &      617.31 &      482.48 &      288.18 \\ \hline
		\end{tabular}
	}
	\label{tab:perf_hw}
\end{table}

\subsection{Hardware and hardware-software performance}
\label{ssec:abs_perf_hw}
Table~\ref{tab:perf_hw} lists the execution times, expressed in milliseconds, of
the considered hardware state-of-the-art implementations of BIKE.
It provides the execution times of the overall KEM as well as a breakdown at the
granularity of KEM primitives for the NIST security level 1 instance of BIKE.

The lightweight, mid-range, and high-performance \textbf{Official} constant-time
implementations~\cite{Richter-Brockmann_TCHES2022} range from 11.14ms to 3.70ms.
The lightweight instance is faster than 64-bit ARM software execution,
while the high-performance one is more than twice faster than
plain-C99 software execution on the Intel CPU but still slower than
the AVX2 software executed on the same Intel CPU, which takes instead 1.06ms.

The \textbf{Client-server} hardware implementation~\cite{Galimberti_DSD2022} improves
over the performance of the official one, with the smaller instance taking 5.74ms
to execute the whole BIKE KEM and the larger one taking instead 0.61ms.
The lightweight instance is thus faster than the official mid-range one,
while the high-performance instance is more than six times faster than the
best-performing official one.
Notably, the authors do not provide a breakdown between the execution times
of the key generation and decapsulation primitives, thus Table~\ref{tab:perf_hw}
provides their aggregate execution time.

Finally, the \textbf{HLS-based} hardware/software solution~\cite{Montanaro_ICECS2022},
which mixes software execution with the adoption of HLS-generated accelerators,
provides an execution time for the overall KEM comprised between 617.31ms and 288.18ms.
While all three instances proposed by the authors outperform
the software execution on the ARM32 CPU, with a speedup up to 2.78$\times$
for the best-performing one, they are however significantly slower
than the software execution on the ARM64 CPU, which takes instead 57.06ms.

The orders of magnitude of difference in the performance between human-designed
hardware implementations and HLS-generated ones highlight the difficulty of
HLS tools to make an efficient use of FPGA resources, in particular for
applications as complex as the BIKE cryptosystem.

\subsection{Hardware and hardware-software resource utilization}
\label{ssec:area_hw}
Table~\ref{tab:area_hw} lists the resource utilization, expressed in terms of
LUT, FF, BRAM, and DSP resources, of the hardware state-of-the-art implementations
of BIKE, and it details the smallest FPGA or SoC that fits the required amount of resources.

\begin{table}[t]
	\centering
	\caption
	[Resource utilization of hardware implementations of BIKE.]
	{Resource utilization of AES-128 security instances of BIKE,
		expressed in terms of LUT, FF, BRAM, and DSP resources,
		for different state-of-the-art FPGA-based implementations.
		Legend: \textbf{LW} lightweight, \textbf{MR} mid-range, \textbf{HP} high-performance instances.}
	\scalebox{1}{
		\begin{tabular}{lrrrrrrrr}
			                  &                                      \multicolumn{8}{c}{\textbf{Hardware implementation}}                                      \\
			                  &  \multicolumn{3}{c}{\textbf{Official}}  & \multicolumn{2}{c}{\textbf{Client-server}} & \multicolumn{3}{c}{\textbf{HLS-based}}  \\
			\textbf{Resource} & \textbf{LW} & \textbf{MR} & \textbf{HP} & \textbf{LW} &                  \textbf{HP} & \textbf{LW} & \textbf{MR} & \textbf{HP} \\ \hline
			\textbf{LUT}      &       12319 &       19607 &       25549 &       51596 &                       217932 &       13567 &       37160 &       50727 \\
			\textbf{FF}       &        3896 &        5008 &        5462 &       29206 &                        97700 &       11621 &       38118 &       49739 \\
			\textbf{BRAM}     &           9 &          17 &          34 &        73.5 &                        632.5 &          40 &          90 &         130 \\
			\textbf{DSP}      &           7 &           9 &          13 &           0 &                            0 &           0 &          35 &          35 \\ \hline
			\textbf{Target}   &       A7-25 &       A7-35 &       A7-50 &      A7-100 &              2$\times$A7-200 &      Z-7010 &      Z-7015 &      Z-7020 \\ \hline
		\end{tabular}
	}
	\label{tab:area_hw}
\end{table}

The \textbf{Official} constant-time implementations~\cite{Richter-Brockmann_TCHES2022}
require the smallest amount of resources, with the lightweight, mid-range, and
high-performance instances fitting respectively on Artix-7 25, 35, and 50 FPGAs.
With respect to the resources available on Artix-7 chips, the most relatively used
resources are LUTs, which thus concur to defining the smallest FPGA which can fit
the BIKE hardware implementation.

The better performance of \textbf{Client-server} implementations~\cite{Galimberti_DSD2022}
comes at the cost of a larger amount of FPGA resources.
In particular, they implement two separate components, one dedicated to
key generation and decapsulation and the other devoted to encapsulation.
The smallest instance proposed by the authors requires an Artix-7 50 chip
for the client core and an Artix-7 35 FPGA for the server one, while the
largest one fits each core on a separate Artix-7 200 chip.
Notably, both the client and server cores do not make use of any DSPs.

Finally, the \textbf{HLS-based} hardware/software instances~\cite{Montanaro_ICECS2022}
target the Zynq-7000 Z-7010, Z-7015, and Z-7020 SoCs.
In particular, the lightweight one implements in hardware the lone key generation primitive,
while the mid-range one implements only decapsulation and the high-performance one
instantiates both the former and latter, resorting to software execution for the lone
encapsulation.
Although not providing a performance that is comparable to the human-designed accelerators,
the HLS-generated accelerators show a significant usage of FPGA resources, in particular
of LUT and BRAM ones.

\section{Conclusions}
\label{sec:concl}
This work provided an overview and a comparison of the software, hardware, and
hardware-software state-of-art implementations of BIKE.

\smallskip
Performance results highlighted significant differences in terms of software execution times
across a variety of computing platforms and software implementations, with the execution
of the whole BIKE KEM taking a time in the order of seconds on lower-end embedded-class ARM CPUs
and a few milliseconds on desktop-class Intel ones with support for AVX2 dedicated instructions.
On the hardware side, the human-designed FPGA-based solutions were shown to outperform the
reference plain-C99 software executed on Intel CPUs.
The best-performing hardware solution could even outperform the AVX2-optimized software,
completing the BIKE execution in 0.61ms compared to the software's 1.06ms.
On the contrary, HLS-generated solutions highlighted the difficulty to generate effective
hardware accelerators through high-level synthesis for target applications as complex as
QC-MDPC code-based cryptosystems.
The considered hardware-software solutions were still able to outperform the reference
software execution on ARM32 CPUs by almost three times.



\bibliography{parma-ditam_2023}

\begin{thebibliography}{10}

\bibitem{BIKE_SW_Github}
{Amazon Web Services - Labs}.
\newblock Additional implementation of bike (bit flipping key encapsulation).
\newblock \url{https://github.com/awslabs/bike-kem}, 2020.

\bibitem{BIKE_website}
Nicolas Aragon, Paulo S. L.~M. Barreto, Slim Bettaieb, Lo\"{i}c Bidoux, Olivier
  Blazy, Jean-Christophe Deneuville, Philippe Gaborit, Shay Gueron, Tim
  G\"{u}neysu, Carlos~Aguilar Melchor, Rafael Misoczki, Edoardo Persichetti,
  Nicolas Sendrier, Jean-Pierre Tillich, Valentin Vasseur, and Gilles
  Z\'{e}mor.
\newblock {BIKE} website.
\newblock \url{https://www.bikesuite.org/}, 2017.

\bibitem{BIKE_spec4.2}
Nicolas Aragon, Paulo S. L.~M. Barreto, Slim Bettaieb, Lo\"{i}c Bidoux, Olivier
  Blazy, Jean-Christophe Deneuville, Philippe Gaborit, Shay Gueron, Tim
  G\"{u}neysu, Carlos~Aguilar Melchor, Rafael Misoczki, Edoardo Persichetti,
  Nicolas Sendrier, Jean-Pierre Tillich, Valentin Vasseur, and Gilles
  Z{\'e}mor.
\newblock {BIKE}: Bit flipping key encapsulation - round 3 submission.
\newblock \url{https://bikesuite.org/files/v4.2/BIKE\_Spec.2021.09.29.1.pdf},
  2021.

\bibitem{LEDAcrypt_website}
Marco Baldi, Alessandro Barenghi, Franco Chiaraluce, Gerardo Pelosi, and Paolo
  Santini.
\newblock {LEDAcrypt} website.
\newblock \url{https://www.ledacrypt.org/}, 2017.

\bibitem{Barenghi_ICECS2019}
Alessandro Barenghi, William Fornaciari, Andrea Galimberti, Gerardo Pelosi, and
  Davide Zoni.
\newblock Evaluating the trade-offs in the hardware design of the ledacrypt
  encryption functions.
\newblock In {\em 2019 26th IEEE International Conference on Electronics,
  Circuits and Systems (ICECS)}, pages 739--742, 2019.
\newblock \href {https://doi.org/10.1109/ICECS46596.2019.8964882}
  {\path{doi:10.1109/ICECS46596.2019.8964882}}.

\bibitem{Bernstein_PKC2006}
Daniel~J. Bernstein.
\newblock Curve25519: New diffie-hellman speed records.
\newblock In Moti Yung, Yevgeniy Dodis, Aggelos Kiayias, and Tal Malkin,
  editors, {\em Public Key Cryptography - PKC 2006}, pages 207--228, Berlin,
  Heidelberg, 2006. Springer Berlin Heidelberg.

\bibitem{Bernstein_Nature2017}
Daniel~J Bernstein and Tanja Lange.
\newblock Post-quantum cryptography.
\newblock {\em Nature}, 549(7671):188--194, 2017.

\bibitem{RacingBIKE_Github}
{Chair for Security Engineering @ Ruhr-Universität Bochum}.
\newblock Racingbike: Improved polynomial multiplication and inversion in
  hardware.
\newblock \url{https://github.com/Chair-for-Security-Engineering/RacingBIKE},
  2021.

\bibitem{Chen_TCHES2021}
Ming-Shing Chen, Tung Chou, and Markus Krausz.
\newblock Optimizing bike for the intel haswell and arm cortex-m4.
\newblock {\em IACR Transactions on Cryptographic Hardware and Embedded
  Systems}, 2021(3):97--124, Jul. 2021.
\newblock URL: \url{https://tches.iacr.org/index.php/TCHES/article/view/8969},
  \href {https://doi.org/10.46586/tches.v2021.i3.97-124}
  {\path{doi:10.46586/tches.v2021.i3.97-124}}.

\bibitem{Chen_ACNS2022}
Ming-Shing Chen, Tim G{\"u}neysu, Markus Krausz, and Jan~Philipp Thoma.
\newblock Carry-less to bike faster.
\newblock In Giuseppe Ateniese and Daniele Venturi, editors, {\em Applied
  Cryptography and Network Security}, pages 833--852, Cham, 2022. Springer
  International Publishing.

\bibitem{Diffie_TIT1976}
W.~Diffie and M.~Hellman.
\newblock New directions in cryptography.
\newblock {\em IEEE Transactions on Information Theory}, 22(6):644--654, 1976.
\newblock \href {https://doi.org/10.1109/TIT.1976.1055638}
  {\path{doi:10.1109/TIT.1976.1055638}}.

\bibitem{Drucker_ARITH2018}
N.~{Drucker}, S.~{Gueron}, and V.~{Krasnov}.
\newblock Fast multiplication of binary polynomials with the forthcoming
  vectorized vpclmulqdq instruction.
\newblock In {\em 2018 IEEE 25th Symposium on Computer Arithmetic (ARITH)},
  pages 115--119, June 2018.
\newblock \href {https://doi.org/10.1109/ARITH.2018.8464777}
  {\path{doi:10.1109/ARITH.2018.8464777}}.

\bibitem{Drucker_CSCML2020}
Nir Drucker, Shay Gueron, and Dusan Kostic.
\newblock Fast polynomial inversion for post quantum qc-mdpc cryptography.
\newblock In Shlomi Dolev, Vladimir Kolesnikov, Sachin Lodha, and Gera Weiss,
  editors, {\em Cyber Security Cryptography and Machine Learning}, pages
  110--127, Cham, 2020. Springer International Publishing.
\newblock \href {https://doi.org/https://doi.org/10.1007/978-3-030-49785-9_8}
  {\path{doi:https://doi.org/10.1007/978-3-030-49785-9_8}}.

\bibitem{Drucker_PQCrypto2020}
Nir Drucker, Shay Gueron, and Dusan Kostic.
\newblock Qc-mdpc decoders with several shades of gray.
\newblock In Jintai Ding and Jean-Pierre Tillich, editors, {\em Post-Quantum
  Cryptography}, pages 35--50, Cham, 2020. Springer International Publishing.

\bibitem{Galimberti_DSD2022}
Andrea Galimberti, Davide Galli, Gabriele Montanaro, William Fornaciari, and
  Davide Zoni.
\newblock Fpga implementation of bike for quantum-resistant tls.
\newblock In {\em 2022 25th Euromicro Conference on Digital System Design
  (DSD)}, 2022.

\bibitem{Galimberti_TC2022}
Andrea Galimberti, Gabriele Montanaro, and Davide Zoni.
\newblock Efficient and scalable fpga design of gf(2m) inversion for
  post-quantum cryptosystems.
\newblock {\em IEEE Transactions on Computers}, 71(12):3295--3307, 2022.
\newblock \href {https://doi.org/10.1109/TC.2022.3149422}
  {\path{doi:10.1109/TC.2022.3149422}}.

\bibitem{Heyse_CHES2013}
Stefan Heyse, Ingo von Maurich, and Tim G{\"u}neysu.
\newblock Smaller keys for code-based cryptography: Qc-mdpc mceliece
  implementations on embedded devices.
\newblock In Guido Bertoni and Jean-S{\'e}bastien Coron, editors, {\em
  Cryptographic Hardware and Embedded Systems - CHES 2013}, pages 273--292,
  Berlin, Heidelberg, 2013. Springer Berlin Heidelberg.

\bibitem{Hu_TC2017}
Jingwei Hu and Ray~C.C. Cheung.
\newblock Area-time efficient computation of niederreiter encryption on qc-mdpc
  codes for embedded hardware.
\newblock {\em IEEE Transactions on Computers}, 66(8):1313--1325, 2017.
\newblock \href {https://doi.org/10.1109/TC.2017.2672984}
  {\path{doi:10.1109/TC.2017.2672984}}.

\bibitem{Hu_TCSII2015}
Jingwei Hu, Wei Guo, Jizeng Wei, and Ray C.~C. Cheung.
\newblock Fast and generic inversion architectures over $\mbox{GF}(2^m)$ using
  modified itoh-tsujii algorithms.
\newblock {\em IEEE Transactions on Circuits and Systems II: Express Briefs},
  62(4):367--371, 2015.
\newblock \href {https://doi.org/10.1109/TCSII.2014.2387612}
  {\path{doi:10.1109/TCSII.2014.2387612}}.

\bibitem{Montanaro_ICECS2022}
Gabriele Montanaro, Andrea Galimberti, Ernesto Colizzi, and Davide Zoni.
\newblock Hardware-software co-design of bike with hls-generated accelerators.
\newblock In {\em 2022 29th IEEE International Conference on Electronics,
  Circuits and Systems (ICECS)}, pages 1--4, 2022.
\newblock \href {https://doi.org/10.1109/ICECS202256217.2022.9970992}
  {\path{doi:10.1109/ICECS202256217.2022.9970992}}.

\bibitem{NIST_website}
{National Institute of Standards and Technology (NIST) - U.S. Department of
  Commerce}.
\newblock Post-quantum cryptography.
\newblock \url{https://csrc.nist.gov/projects/post-quantum-cryptography}, 2021.

\bibitem{NIST_IR8413}
{National Institute of Standards and Technology (NIST) - U.S. Department of
  Commerce}.
\newblock Nistir 8413, status report on the third round of the nist
  post-quantum cryptography standardization process.
\newblock \url{https://nvlpubs.nist.gov/nistpubs/ir/2022/NIST.IR.8413.pdf},
  2022.
\newblock \href {https://doi.org/https://doi.org/10.6028/NIST.IR.8413}
  {\path{doi:https://doi.org/10.6028/NIST.IR.8413}}.

\bibitem{NSA_CNSA2.0}
{National Security Agency}.
\newblock Commercial national security algorithm suite 2.0 (cnsa 2.0)
  cybersecurity advisory (csa).
\newblock
  \url{https://media.defense.gov/2022/Sep/07/2003071834/-1/-1/0/CSA_CNSA_2.0_ALGORITHMS_.PDF},
  2022.

\bibitem{Niederreiter_PCIT1986}
Harald Niederreiter.
\newblock Knapsack-type cryptosystems and algebraic coding theory.
\newblock {\em Prob. Contr. Inform. Theory}, 15(2):157--166, 1986.

\bibitem{Richter-Brockmann_TCHES2022}
Jan Richter-Brockmann, Ming-Shing Chen, Santosh Ghosh, and Tim G\"{u}neysu.
\newblock Racing bike: Improved polynomial multiplication and inversion in
  hardware.
\newblock Cryptology ePrint Archive, Paper 2021/1344, 2021.
\newblock \url{https://eprint.iacr.org/2021/1344}.
\newblock URL: \url{https://eprint.iacr.org/2021/1344}.

\bibitem{Richter-Brockmann_TC2021}
Jan Richter-Brockmann, Johannes Mono, and Tim G\"{u}neysu.
\newblock Folding bike: Scalable hardware implementation for reconfigurable
  devices.
\newblock {\em IEEE Transactions on Computers}, 2021.
\newblock \href {https://doi.org/10.1109/TC.2021.3078294}
  {\path{doi:10.1109/TC.2021.3078294}}.

\bibitem{Rivest_CommACM1978}
R.~L. Rivest, A.~Shamir, and L.~Adleman.
\newblock A method for obtaining digital signatures and public-key
  cryptosystems.
\newblock {\em Commun. ACM}, 21(2):120--126, February 1978.
\newblock \href {https://doi.org/10.1145/359340.359342}
  {\path{doi:10.1145/359340.359342}}.

\bibitem{Shoup_IACR2001}
Victor Shoup.
\newblock A proposal for an iso standard for public key encryption.
\newblock Cryptology ePrint Archive, Paper 2001/112, 2001.
\newblock \url{https://eprint.iacr.org/2001/112}.
\newblock URL: \url{https://eprint.iacr.org/2001/112}.

\bibitem{Maurich_DATE2014}
Ingo von Maurich and Tim G\"{u}neysu.
\newblock Lightweight code-based cryptography: Qc-mdpc mceliece encryption on
  reconfigurable devices.
\newblock In {\em 2014 Design, Automation and Test in Europe Conference \&
  Exhibition (DATE)}, pages 1--6, 2014.
\newblock \href {https://doi.org/10.7873/DATE.2014.051}
  {\path{doi:10.7873/DATE.2014.051}}.

\bibitem{Zoni_Access2020_Dec}
D.~{Zoni}, A.~{Galimberti}, and W.~{Fornaciari}.
\newblock Efficient and scalable fpga-oriented design of qc-ldpc bit-flipping
  decoders for post-quantum cryptography.
\newblock {\em IEEE Access}, 8:163419--163433, 2020.
\newblock \href {https://doi.org/10.1109/ACCESS.2020.3020262}
  {\path{doi:10.1109/ACCESS.2020.3020262}}.

\bibitem{Zoni_Access2020_Mul}
D.~{Zoni}, A.~{Galimberti}, and W.~{Fornaciari}.
\newblock Flexible and scalable fpga-oriented design of multipliers for large
  binary polynomials.
\newblock {\em IEEE Access}, 8:75809--75821, 2020.
\newblock \href {https://doi.org/10.1109/ACCESS.2020.2989423}
  {\path{doi:10.1109/ACCESS.2020.2989423}}.

\end{thebibliography}

\end{document}